\documentclass[10pt,conference]{IEEEtran} 

\usepackage[utf8]{inputenc}
\usepackage[T1]{fontenc}
\usepackage{microtype}
\usepackage{graphicx}
\usepackage{nccmath}
\usepackage{xcolor}
\usepackage{newtxmath}
\usepackage{physics}
\usepackage{xfrac} 
\usepackage{siunitx}
\usepackage{mathtools}
\usepackage{hyperref}
\usepackage{float}
\usepackage{cite}
\usepackage[caption=false,font=footnotesize]{subfig}
\usepackage[textsize=tiny]{todonotes}

\setuptodonotes{fancyline}
\newcommand{\fred}[2][]{\todo[color=green!30, #1]{\textbf{FG:} #2}}  

\newcommand{\trt}{t_\rightleftharpoons} 

\makeatletter
\tikzstyle{notestyleraw} = [
draw=\@todonotes@currentbordercolor,
fill=\@todonotes@currentbackgroundcolor,
text=\@todonotes@currenttextcolor,
line width=0.5pt,
text width = \@todonotes@textwidth - 1ex - 1pt, 
inner sep = 0.4 ex,
rounded corners=4pt,
yshift=0cm]
\makeatother

\begin{document}

\title{Entanglement Swapping in Orbit:\\ 
a Satellite Quantum Link Case Study}
\author{%
    \IEEEauthorblockN{%
        Paolo Fittipaldi\IEEEauthorrefmark{1}, 
        Kentaro Teramoto\IEEEauthorrefmark{2}, 
        Naphan Benchasattabuse\IEEEauthorrefmark{3}\IEEEauthorrefmark{5},\\
        Michal Hajdušek\IEEEauthorrefmark{3}\IEEEauthorrefmark{5}, 
        Rodney Van Meter\IEEEauthorrefmark{4}\IEEEauthorrefmark{5}, 
        Frédéric Grosshans\IEEEauthorrefmark{1}
        }
    \IEEEauthorblockA{
        \IEEEauthorrefmark{1}Sorbonne Université,  CNRS, LIP6, F-75005 Paris, France \\           
    }
    \IEEEauthorblockA{
        \IEEEauthorrefmark{2}Mercari R4D, Mercari, Inc., Japan
    }
    \IEEEauthorblockA{
        \IEEEauthorrefmark{3}Graduate School of Media and Governance, Keio University Shonan Fujisawa Campus, Kanagawa, Japan
    }
    \IEEEauthorblockA{
        \IEEEauthorrefmark{4}Faculty of Environment and Information Studies, Keio University Shonan Fujisawa Campus, Kanagawa, Japan
    }
    \IEEEauthorblockA{
        \IEEEauthorrefmark{5}Quantum Computing Center, Keio University, Kanagawa, Japan
    }
    \IEEEauthorblockA{\{paolo.fittipaldi,frederic.grosshans\}@lip6.fr, zigen@mercari.com, \{whit3z,michal,rdv\}@sfc.wide.ad.jp}
}

\maketitle
\begin{abstract}
Satellite quantum communication is a promising way to build long distance quantum links, making it an essential complement to optical fiber for quantum internetworking beyond metropolitan scales. 
A satellite point to point optical link differs from the more common fiber links in many ways, both quantitative
(higher latency, strong losses) and qualitative (nonconstant parameter values during satellite passage, intermittency of the link,
impossibility to set repeaters between the satellite and the ground station).
We study here the performance of a quantum link between two ground stations, 
using a quantum-memory-equipped satellite as a quantum repeater. 
In contrast with quantum key distribution satellite links,
the number of available quantum memory slots m, together with the unavoidable round-trip communication latency t of at least a few
milliseconds, severely reduces the effective average repetition rate to m/t --- at most a few kilohertz for foreseeable quantum memories. 
Our study uses two approaches, which validate each other: 1) a simple analytical model of the effective rate of the quantum link;
2) an event-based simulation using the open source Quantum Internet Simulation Package (QuISP).
The important differences between satellite and fiber links led us to modify QuISP itself.
This work paves the way to the study of hybrid satellite- and fiber-based quantum repeater networks interconnecting
different metropolitan areas.
\end{abstract}

\section{Introduction}

\IEEEPARstart{S}{atellite-based} optical links have been shown\cite{dFdP,YinMicius,Micius} to support the distribution of entangled pairs of qubits over continental distances. 
An important potential application of quantum satellite link studies is the interconnection of metropolitan-scale fiber quantum subnetworks \cite{AwschalomInterconnects,RajaConnectingCities}: 
developing such a hybrid architecture would pave the way to the construction of a global quantum internet capable of distributing quantum entanglement to arbitrary groups of users, making worldwide distributed quantum applications\cite{protocolzoopaper,protocolzoo,AwschalomRoadmap} possible.

In order to better understand this elementary brick, we show here a detailed 
study of a quantum link between Paris and Nice which could be created through
a single passage of a quantum-memory equipped satellite following the orbit of Micius \cite{Micius}.
We approach the problem through two complementary points of view: a simple 
analytical model and an event-based simulation.

Given that ground--satellite communication happens on a millisecond time scale, during which the satellite does not move appreciably, a notion of rate of the link at a given time makes sense, allowing
us to derive a set of simple analytical formulas depending on known link parameters through which we can estimate the performance of a given satellite link. 
This model provides valuable insight during the feasibility study and network design phases, while also posing as a validation tool for results obtained through other means.

On the simulation side, we developed new components that we integrated into the open source Quantum Internet Simulation Package (QuISP)\cite{quisp,wwwquisp}, an event-based quantum network simulator that focuses on scalability and protocol design while retaining physical accuracy and high configurability. 
Beyond the evaluation of the performance of the link itself, the scalability of
QuISP should allow us to simulate the integration of a satellite link
between several terrestrial quantum networks.

This paper deals with the analysis of the performance of a single quantum satellite link in the satellite--ground and ground--satellite--ground scenarios.
For simplicity reasons, we start with the first case, even if the second is more relevant.
In the first case, we study entanglement distribution between a quantum-memory-equipped satellite
following Micius' orbit \cite{Micius} and a ground station in Nice, in the south of France.
We assume a given number $m_S$ of quantum memory slots
(both onboard the satellite and at the ground station) and demonstrate the impact of memory 
size on the performance of our link.

In the second case, we analyze a dual link between the satellite and two ground stations, 
one in Nice and one in Paris, with the satellite node acting as a quantum repeater. 
Once link-level entanglement is established between the satellite and each of the stations, 
the satellite creates a direct Nice--Paris connection through entanglement swapping. 

Our goal is to provide a simple model and a simulation interface that are modular enough to be of use in general quantum internet research: our analytical model's simplicity makes it 
useful for 
large network design calculations, while our simulation modules are seamlessly integrated in a realistic and highly scalable quantum network simulator, enabling simulation of large hybrid networks including satellite links.

We start our paper with Sec.\@ \ref{sec:LitReview}, a contextualization of our work in the literature. Sec.\@ \ref{sec:SysDesc} describes the system under investigation and our assumptions. In Sec.\@ \ref{sec:TheorySingleLink}, we describe our theoretical model for a single satellite-to-ground link and cross-validate it with simulation data. Sec.\@  \ref{sec:TheoryDualLink} extends our model to swapping-based dual satellite-to-ground links, which are the ones that bear communication relevance. In this section, we also discuss the problem of allocating memory slots aboard the satellite. 
Sec.\@ \ref{sec:SoftwareStack} is a brief overview of the software stack, our motivations behind the choice of QuISP and the modifications we applied to the code base to support satellite links.
Finally, Sec.\@ \ref{sec:conclusion} concludes the paper, highlights the shortcomings of this work and proposes future research directions. 

\section{Context and Relevance of this Work}
\label{sec:LitReview}
In this section, we review a selection of interesting papers connected to our contribution, which is summarized at the end.

In \cite{dFdP}, de Forges de Parny et al.\@ provide a comprehensive review of satellite quantum communication, identifying the main technical issues and providing an in-depth discussion and state-of-the-art for all the involved subsystems and components. A simulation section is also provided for a satellite link connecting Paris and Nice that differs from the present contribution in two key ways: first, \cite{dFdP} focuses on memoryless quantum links, neglecting the impact of memory on the link, which is one of the key investigation directions of this work. Second, the referenced paper employs a purpose-built simulation tool to present feasibility results, whereas one of the main focal points of the present work is the extension of a pre-existing simulation package in order to retain maximal interoperability and extensibility.

In \cite{Micius}, Lu et al.\@ deliver a complete report of the design of a real satellite system (the Micius mission), shedding light on the practical solutions to several experimental challenges in designing and operating a quantum satellite.  

Yehia et al.\@ \cite{RajaQuantumCity,RajaConnectingCities} share a similar end goal with this work: in the first cited paper, they describe and simulate the architecture of a ``quantum city'', i.e.\@ a metropolitan-scale quantum network, while in the second one the connection of quantum cities through satellite links is discussed. With respect to this work, Yehia et al.\@ focus more on the simulation of quantum key distribution (QKD) across a European quantum network through NetSquid\cite{netsquid}, a more physics-oriented simulator that, albeit more accurate and versatile than QuISP in terms of physical backend, is less focused on networking, realistic modeling of control-related communication and protocol design. 
In contrast with Yehia et al.\@, we do not discuss QKD in this work, preferring to focus instead on the entanglement distribution aspect of the satellite link under a generic application and its inclusion in a general internetworking context. 

Ref.\@  \cite{QInternetArchitecture,quisp}, which share several authors with the present work,
are closely related to our contribution,
in that they give a precise picture of QuISP and the quantum network architecture it simulates: the former paper outlines a recursive network architecture for quantum networks enabling the setup of arbitrary connections that implement generic applications, 
while the latter is a detailed introduction to QuISP, describing the key design decisions, demonstrating its performance and scalability and generally providing users with a starting point for quantum network simulation.

Our work's contribution to the satellite communication field is essentially twofold:\begin{itemize}
    \item We provide, to our knowledge, the first analytical investigation of a memory-equipped satellite quantum link and examine it in terms of realistic design questions such as memory size and classical communication latency;
    \item We simulate the satellite link employing a pre-existing quantum communication simulator focused on scalability and extensibility, our ultimate goal being the development of a building block that can be placed inside a planet-scale network topology for the simulation of arbitrary quantum applications at large scales. 
\end{itemize}
\section{System Description}
\label{sec:SysDesc}
\begin{figure}
    \centering
    \includegraphics[width=\linewidth]{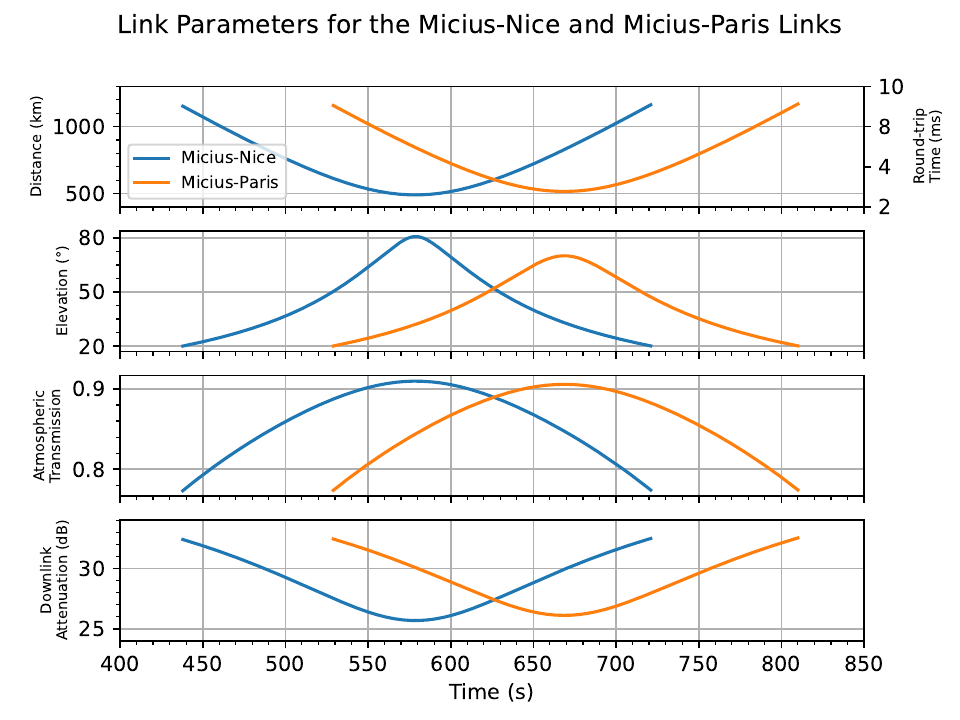}
    \caption{Parameters of both the Micius--Nice and the Micius--Paris links during the passage of the Micius satellite
        starting January 2$^{\text{nd}}$, 2023 at 10:00 AM. 
        The distance and elevation  are computed using \texttt{orekit} \cite{orekit} and the atmospheric attenuation with \texttt{LOWTRAN} \cite{LOWTRAN}, in a rural setting with $23 \unit{km}$ visibility  and 
        no aerosol at a wavelength of $1550\,\unit{nm}$.
        The channel attenuation is computed using the previous parameters, using 
        (1) of \cite{SatLosses} for a satellite-to-ground (downlink) transmission
        with the parameters of \cite{dFdP}, 
        assuming telescope diameters of 10~cm on the satellite and a 1~m diameter on the ground. The link is supposed to be available only for elevations above 20°.
    \label{fig:LinkParams}
    }
\end{figure}

The link we study operates between two ground stations located in Nice and Paris, 
two French cities approximately 680 km apart. 
Their large difference in latitude explains the $91 \unit{s}$ time difference in closest approach of the satellite from both locations, 
a nonnegligible difference for a 5 minute long satellite passage, while their longitude difference explains the $\ang{11}$
difference in their maximal elevation.
As a result, the link budget varies independently along the two legs of a ground-satellite-ground link. To keep our discussion realistic, we employ orbital data from
a single passage of the Micius satellite \cite{Micius}. 
The resulting parameters are detailed in Fig.\@ \ref{fig:LinkParams}.

This study is in the context of first generation quantum repeater networks\cite{1g2g3g}, 
where each node ---both on the ground and in space--- has access to a quantum memory, 
which it uses to synchronize entanglement swapping.
A key parameter of such memory is its lifetime $\tau$, the timescale at which
decoherence occurs. 
Given the unavoidable $\trt \ge 2\unit{ms}$ satellite--to--ground roundtrip communication time,
any quantum memory where $\tau \ll 1\unit{ms}$ would be essentially useless for 
our purpose. At the opposite end, quantum memories with very long
lifetimes ---$\tau\ge 100 \unit{s}$--- would allow sneakernet-like\cite{rfc1149} 
protocols where the satellite catches some qubits to carry them elsewhere, as in \cite{QNetShip}. 
To simplify our analysis, we set ourselves in an intermediate regime, 
where $\trt\ll\tau < 10 \unit{s}$: the decoherence
of the memory is negligible, but the satellite either uses or swaps the qubits within a few
seconds ---actually most often a few tens of milliseconds--- a timescale practically
instantaneous compared to its movement.
The very relevant $\tau\sim 1\unit{ms}$ regime, where the decoherence of the quantum memory
induces noise, is kept for future work.

A limitation of the quantum memory we take into account is its finite storage capacity
$m$. In practice, the limiting factor will be the number of qubits $m_S$ the satellite
can store, which will be taken to be $m_S\in\qty{10,50,100}$ for numerical applications.
The satellite is able to emit a photon entangled with a qubit of its memory, as well as
to perform a perfect Bell measurement between two qubits stored inside it.

At each ground station, we assume the same hardware as the satellite plus a Bell State Analyzer, which is required by the link-level entanglement generation protocol. Throughout the paper, we rely on the \texttt{SenderReceiver} protocol  described in \cite{Jones} for establishment of link-level quantum entanglement. In short, the protocol requires one of the nodes establishing entanglement (the Sender) to generate a stream of photons entangled with local memory slots. The stream of photons is to be relayed to the Receiver, which performs interference measurements in order to ``latch'' the incoming qubits in a local quantum memory. Entanglement generation through this protocol works in rounds of $N$ attempts, where $N$ is the minimum number of available memory slots at the two nodes. Once $N$ attempts have been made, the Receiver shares the measurement results with the Sender, failed entanglement attempts are cleared from memory and a new round begins. 

\section{Satellite--Ground Link Analysis}
\label{sec:TheorySingleLink}
It is already apparent from the high-level description of the protocol we adopt how classical communication latency plays a crucial role in the calculation of the entanglement distribution rate: an entanglement generation round requires one round trip of the channel 
---milliseconds for a satellite link--- effectively stunting the attainable entanglement distribution rate. 

\subsection{Entanglement Rate}
In particular, let $m_G$ and $m_S$ be the number of memory slots available respectively at the ground station and onboard the satellite. We assume $m_S\le m_G$, for technological reasons.
Let $\eta$ be the downlink transmission coefficient of the satellite to ground channel calculated in Fig.\@ \ref{fig:LinkParams}.
The generation of a satellite-ground entangled pair needs the storage of one
qubit in the satellite quantum memory for a roundtrip time $\trt$ ---until confirmation 
of the reception (or lack thereof) of the photon by the ground station is received---
and is successful with a probability $\eta$. 
Thus, taking $N = m_S$ and letting $p_{\text{BSM}}$ be the success probability of the latching Bell State Measurement ($\sfrac{1}{2}$ in our case \cite{weinfurter1994experimental}), the best attainable entanglement distribution rate will be
\begin{align}
\label{eq:theoretical-rate-uncorrected}
    r \leq p_{\text{BSM}}\frac{\eta N}{\trt}.
\end{align}
Although (\ref{eq:theoretical-rate-uncorrected}) is already surprisingly accurate given its
simplicity (as shown later in Fig.\@ \ref{fig:validation-matchrate} and \ref{fig:validation-matchnumber}), 
some discrepancy appears for high $m_S=100$. 
As discussed below, this is explained by the differential latency induced by the satellite speed.

\subsection{Differential Latency}
\begin{figure}
    \centering
    \includegraphics[width=\linewidth]{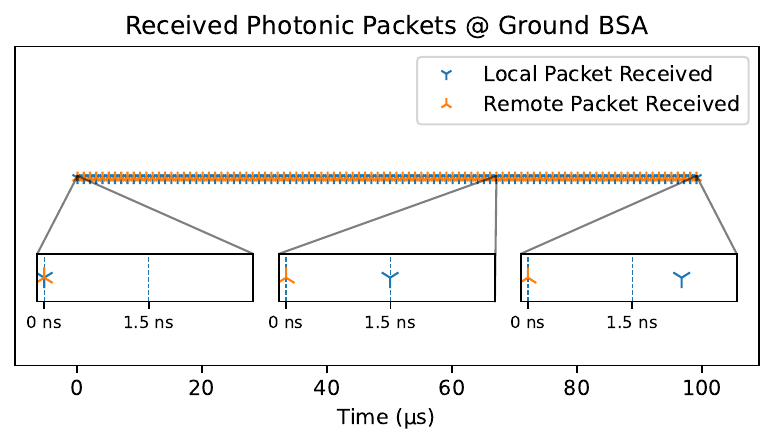}
    \caption{Timing of the first photon train, for a Micius--Nice
        link with $m_S=100$, extracted from our simulation. The reception of the first packet sets $t=0$, and the repetition rate is 1 \unit{MHz}. The local packet corresponds to
        the photon emitted by the ground quantum memory and the remote one to the photon received
        from the satellite. As shown by the inset, while initially in sync, they progressively 
        desynchronize, the 67$^\text{th}$ being the last one inside the acceptance window. A value of 1.5 \unit{ns} (default in QuISP) was employed for the acceptance window because it is technologically reasonable and consistent with our other parameters.\label{fig:difflatency_demo}
    }
\end{figure}
Keeping in mind how crucial latency is to the efficient generation of entangled pairs, 
the high speed of satellites ---a few km/s--- introduces additional complexity in the form of \textit{differential latency},
that is a change of the timing of arrival of the photons due to the movement of the emitting satellite.
The \texttt{SenderReceiver} protocol requires the Receiver to interfere incoming photons with local ones, 
implying that the changing timing of incoming photons should be taken into account throughout the procedure.
In theory, nothing prevents perfect compensation of the latency offsets, but practical considerations can make it difficult.
In our simulation, we have assumed classical communications to allow a perfect synchronization of the first photon of
each train,
but have not corrected for the difference in the apparent photon period.
This implies that successive photons in the same train will eventually drift out of sync (as shown in Fig.\@ \ref{fig:difflatency_demo}), introducing an interplay between emission frequency, acceptance window for the interferometric measurement and satellite velocity. 

Due to differential latency, a time discrepancy is created that accumulates over successive photons until two corresponding photons do not interfere anymore and latching fails. Letting $\delta t_{DL}$ be the time shift introduced by the satellite's displacement, $v_r$ the radial component of the satellite's velocity, $T_{em}=1\unit{\micro s}$ the period of photon emission and $c=3\cdot10^8 \unit{m/s}$ the speed of light, we have
\begin{align}
    \delta t_{DL} = \frac{v_r T_{em}}{c}.
\end{align}
Since each successive photon undergoes cumulative desyncing, if we assume $v_r$ to be constant over the duration $m_ST_{em}$ of
one photon train we can place an upper bound on the useful photon train length $N$:
\begin{align}
\label{eq:maxN}
    N \leq \frac{w_i c}{|v_r| T_{em}},
\end{align}
where $w_i$ 
corresponds to the acceptance window of the Bell State Measurement at the ground station (1.5 \unit{ns} in our case, as shown in fig. \ref{fig:difflatency_demo}). Substituting (\ref{eq:maxN}) inside (\ref{eq:theoretical-rate-uncorrected}) yields a new expression for the theoretical rate, corrected for differential latency:
\begin{align}
\label{eq:theoretical-rate-corrected}
     r^* = p_{\text{BSM}}\frac{\eta}{\trt}\min\left(m_S,\frac{w_i c}{|v_r| T_{em}}\right).
\end{align}
This expression demonstrates how not only the classical communication latency drastically lowers entanglement distribution rates, but the number of photons that is useful to exchange in a single experimental round is linked to the satellite's velocity, further impairing performance in the faraway orbit sections. 
\subsection{Validation of the Single Link Scenario}

\begin{figure}
    \centering
    \includegraphics[width=\linewidth]{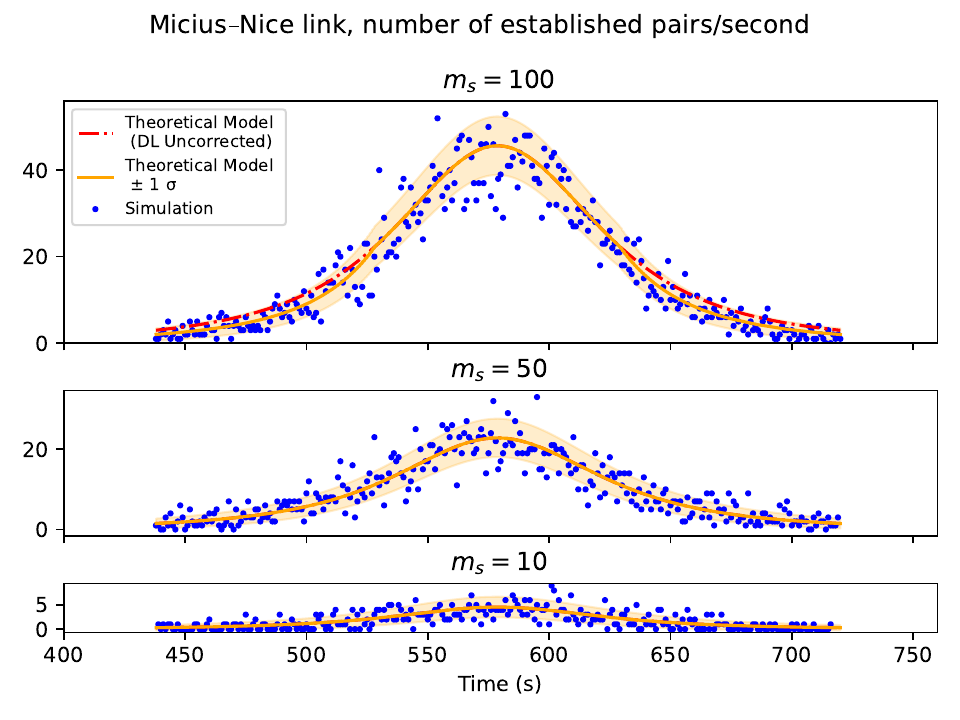}
    \caption{Rate of entanglement generation over a single Micius--Nice link for quantum memory sizes 
        $m_S\in\{10,50,100\}$.
        The orange shaded area corresponds to the expected $1\sigma$ statistical fluctuations.
        The simulation datapoints (blue points) for each second are within expected fluctuations.}
    \label{fig:validation-matchrate}
\end{figure}
\begin{figure}
    \centering
    \includegraphics[width=\linewidth]{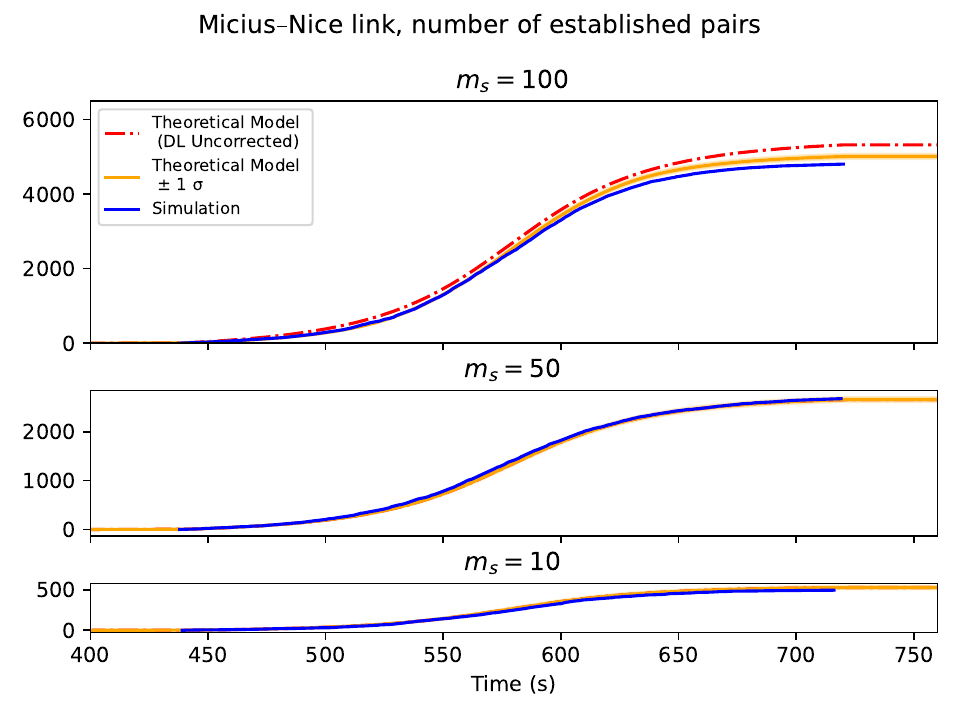}
    \caption{Number of generated entangled pairs over a single Micius--Nice link for quantum memory $m_S\in\{10,50,100\}$.
    The orange shaded area corresponds to the expected $1\sigma$ statistical fluctuations.
    The simulation (blue curve) is within expected fluctuations.
    \label{fig:validation-matchnumber}}
\end{figure}
In this subsection, we provide results that cross-validate our model and simulation. We ran our simulations for a single passage 
of the Micius\cite{Micius} satellite, as seen by the Nice ground station for different entanglement round lengths (as detailed in 
Fig.\@ \ref{fig:validation-matchrate} and \ref{fig:validation-matchnumber}) and observed satisfactory agreement with the theoretical
calculations both in terms of entanglement distribution rate and total number of entangled pairs.  
Additionally, we provide a plot of the predicted uncorrected rate of incoming photons $r$ (as per (\ref{eq:theoretical-rate-uncorrected})). 
For the maximal radial velocity of the satellite compared to Nice which is 6998~\unit{m/s}, 
(\ref{eq:maxN}) becomes $N\le 67$, which explains why the correction is only relevant for $m_S=100$, but not for $m_S\in\qty{10, 50}$.
A close look at Fig. \ref{fig:validation-matchrate} further shows that the correction is relevant at the beginning and end of the passage,
when the satellite is further away from the zenith
and its radial velocity higher.

\section{Ground--Satellite--Ground Link Analysis}
\label{sec:TheoryDualLink}
\begin{figure*}
\centering
\includegraphics[width=\textwidth]{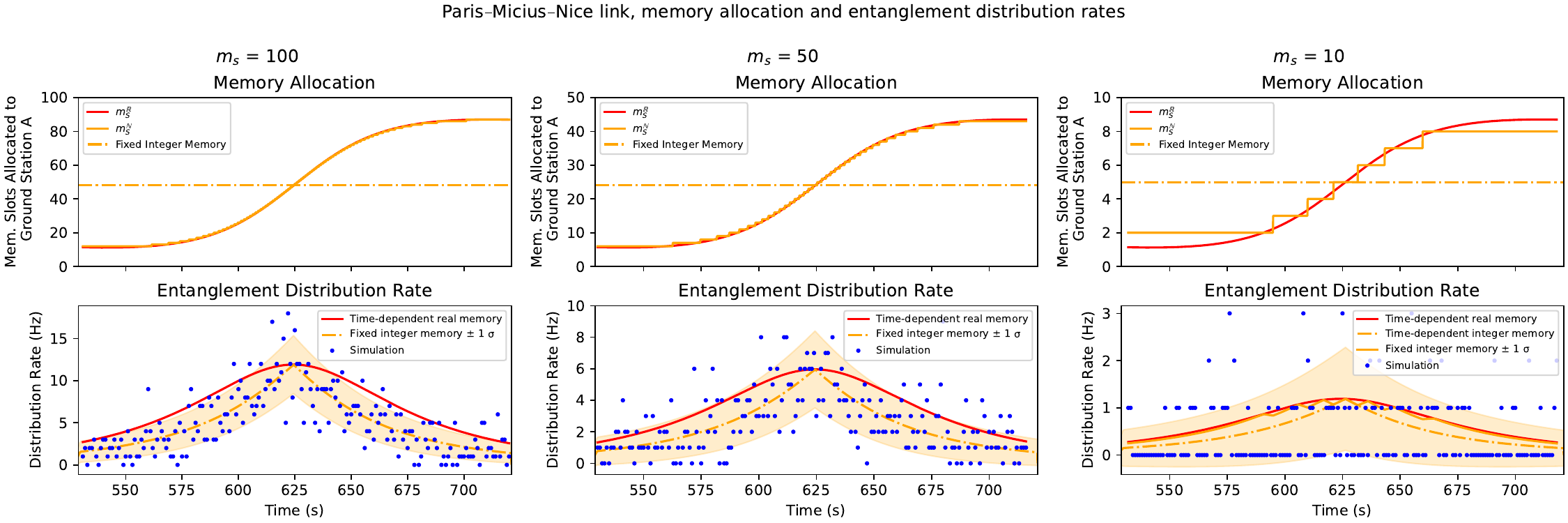}
\caption{\label{fig:malloc}
Comparison of the theoretical entanglement distribution rate over a swapping Nice--Micius--Paris link with dynamic memory allocation against the theoretical and simulated 
    distribution rates with fixed allocation.
    The simulation datapoints (blue points) for each second are within expected fluctuations.}
\end{figure*}


\begin{figure}
    \centering
    \includegraphics[width=\linewidth]{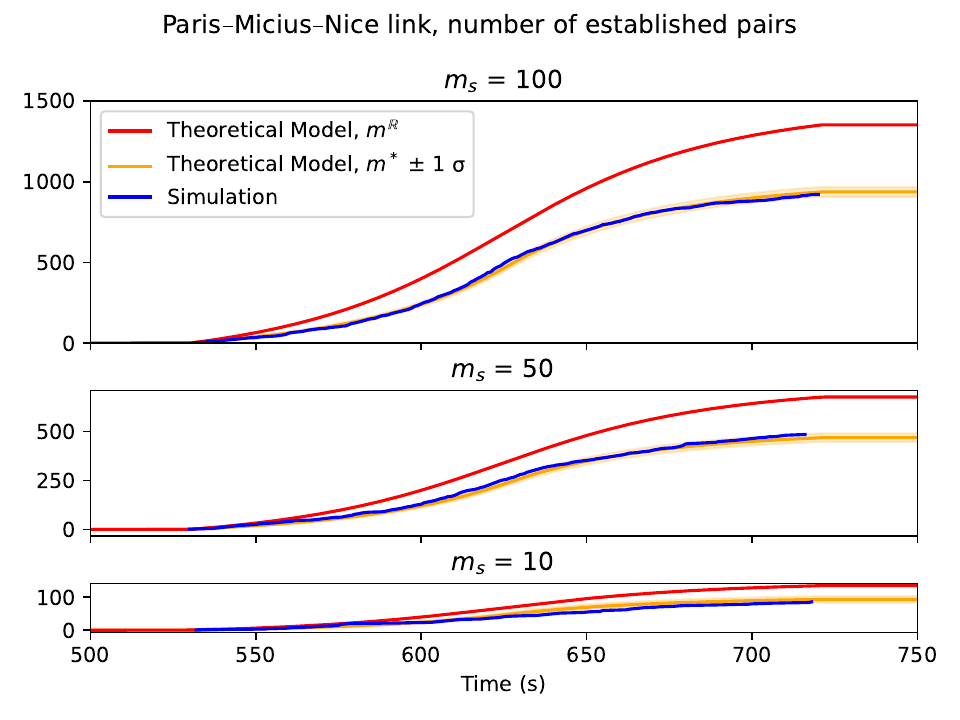}
    \caption{Number of generated entangled pairs over a swapping Nice--Micius--Paris link for 
    quantum memory sizes $m_S\in\{10,50,100\}$.
     The performance difference between static and dynamic allocation is of 36\%, showing the interest of the latter for future work, even if both strategies do not differ by orders of magnitude.
     The simulation results are within expected fluctuations.}
    \label{fig:validation-twinmatchrate}
\end{figure}
To analyze the dual link case, we adopt once again the \texttt{SenderReceiver} protocol to generate entanglement between the satellite and each of the ground nodes, and subsequently perform entanglement swapping operations aboard the satellite, thus merging the two satellite-to-ground links.
The dual entanglement distribution rate is simply the minimum between the entanglement distribution rates of each leg of the system,
computed as previously shown. However, a question remains: given a fixed amount $m_S$ of available memory slots aboard the satellite, 
how should they be assigned to each link?

Since the parameters of both downlinks evolve differently, there is no reason for
 a static $50:50$ split of the available slots to be the best solution. 
 In the following, we start by studying optimal dynamic memory allocation. 
 However, for various reasons, a constant split is simpler (for instance, when working with separate network interfaces connected to distinct quantum memories, as per \cite{QInternetArchitecture}). 
 Thus, we will work with a fixed value, extracted from the preceding dynamical analysis.
We use the obtained values to validate the theoretical model against simulation and assess the performance degradation introduced by adopting static vs.\@ dynamic memory allocations. 

Let us first assume the number of allocated memory slots to each link, $m_A$ and $m_B$, can take real (non-integer) time-dependent values, 
where $A$ and $B$ label the two satellite--ground downlinks, and $m_A+m_B = m_S$. 
The entanglement distribution rate $r_{AB}$ between the two ground stations is equal to the minimum of the two single-link rates.
Using (\ref{eq:theoretical-rate-uncorrected}), neglecting the differential latency correction, we have
\begin{align}
\label{eq:twinrate}
    r_{AB} = \min\left(p_{\text{BSM}}^A\frac{m_S^A}{\trt^A}\eta^A,p_{\text{BSM}}^B\frac{ m_S^B}{\trt^B}\eta^B \right),
\end{align}
where 
$\trt^\chi$ is the round-trip time across channel $\chi\in \qty{A,B}$, $\eta^\chi$ the corresponding transmission coefficient and $p_{\text{BSM}}^\chi$ the success probability of the latching Bell State Measurement at the corresponding ground station.
This expression is maximized when both arguments of the $\min$ are equal, leading to
\begin{align}
\label{eq:memory_repartition}
    m^{\mathbb{R}}_A(t) &= \frac{\sfrac{\trt^A}{\eta^A}}{\sfrac{\trt^A}{\eta^A} + \sfrac{\trt^B}{\eta^B}}m_S,&
    m^{\mathbb{R}}_B(t) &= \frac{\sfrac{\trt^B}{\eta^B}}{\sfrac{\trt^A}{\eta^A} + \sfrac{\trt^B}{\eta^B}}m_S.
\end{align}
Note that all quantities except for $m_S$ are time-dependent. 

Substituting the relevant channel data, we obtain a time-dependent, real expression for the memory allocated to either link.
Of course, any actual memory partition can only accept integers value $m^{\mathbb{N}}_\chi(t)$,
optimally given by
\begin{align}
\label{eq:Nmemory_repartition}
    m^{\mathbb{N}}_A &= \left\lceil\frac{(m_S-1)\sfrac{\trt^A}{\eta^A}}{\sfrac{\trt^A}{\eta^A} + \sfrac{\trt^B}{\eta^B}}\right\rceil,&
    m^{\mathbb{N}}_B &=  \left\lfloor\frac{(m_S-1)\sfrac{\trt^B}{\eta^B}}{\sfrac{\trt^A}{\eta^A} + \sfrac{\trt^B}{\eta^B}} +1\right\rfloor.
\end{align}
As shown in Fig.\@ \ref{fig:validation-twinmatchrate}, even for $m_S=10$, this integer allocation
leads to rates close to the ones obtained using (\ref{eq:memory_repartition}).

To simplify the simulation, and because QuISP currently does not allow dynamical memory allocations between different
“Quantum Network Interface Cards” (QNICs), 
we have restricted ourselves to fixed memory allocation, taking as fixed value $m^*_A$ 
the memory value $m_A^\mathbb{N}(t)$  for which  $r_{AB}(m_A^\mathbb{N}(t))$ attains its maximal value.

This  yields $(m^*_A,m^*_B)$, fixed integer memory allocation values that we can both substitute in (\ref{eq:twinrate}) to obtain a theoretical rate and input to our simulation as memory slot numbers.
 From following the procedure above for the case of $m_S\in\{ 100,50,10\}$ with link $A$ associated to the Micius--Nice leg and link $B$ to the Micius--Paris one we obtain memory splits $(m^*_A,m^*_B) = (48,52), (24,26)$ and $(5,5)$ respectively. 
 Fig.\@ \ref{fig:malloc} and \ref{fig:validation-twinmatchrate} show a comparison between the different allocation strategies, together
 with the simulation results for a fixed memory split.

\section{Simulation Using QuISP}
\label{sec:SoftwareStack}
To perform our simulations, we have modified the Quantum Internet Simulation Package (QuISP)\cite{quisp,wwwquisp}, which is an event-driven simulator based on OMNeT++\cite{omnetpp,wwwomnetpp}, a well-established simulation platform 
for classical networks. 
In this section, we provide a brief overview of OMNeT++ and QuISP to motivate the choice of our software stack
and describe our work to integrate dynamic satellite links into this tool targeted towards more static fiber quantum networks.

\subsection{OMNeT++ and QuISP}
Every simulation using OMNeT++ is defined through \textit{modules}, 
i.e.\@ basic functional units that execute arbitrary C++ code, and \textit{messages} for the modules to exchange. 
Every time a module receives a message, it executes a given portion of its internal code to simulate networking.  Such a simple basic structure has the advantage of high scalability and fidelity to the simulated network system: since everything happens through messages, the simulation mirrors the internal workings of a network system on a message-by-message basis. Moreover, since the channels that interconnect modules are themselves customizable, it is possible to implement and study simulated networks that closely reproduce the physical transfer of signals. 

Built on top of OMNeT++, QuISP provides a suite of modules and channels that were designed for the simulation of quantum fiber networks. In short, the control architecture implemented by QuISP revolves around the RuleSet\cite{MatsuoRuleset,CocoriMaster,QInternetArchitecture}, a list of arbitrary instructions that can be executed by the processing modules inside the quantum nodes to implement general quantum applications. While such an abstract architecture has numerous advantages, the most interesting one in this context is the decoupling between the quantum hardware, the classical communication components and the control modules: since a quantum node acts as a self-contained processor for arbitrary RuleSets, it is possible to implement a new interface that extends the quantum node's communication and interfacing capabilities while leaving the quantum hardware and its control logic virtually untouched, retaining the possibility to interoperate the modified satellite components with the original fiber ones in a large-scale hybrid network system. 
A detailed description of QuISP can be found in \cite{quisp,wwwquisp}.

\subsection{Adapting QuISP to Dynamic Satellite Networks}
In this section, we summarize the challenges encountered during the adaptation of QuISP to satellite systems.
The code changes corresponding to this work are available in \cite{PR} and should be soon integrated 
into the main branch of QuISP.

Since QuISP was designed with wired fiber networks in mind, several assumptions ingrained in the code base required modifications and extensions to allow adaptation to the more dynamical nature
of satellite communications.
All our changes can be summed up as allowing the variation of parameters
which are fixed in fiber networks, and managing the consequences.

The two main areas where QuISP's code base was extended were in the channel suite, where a new \texttt{FreeSpaceChannel} object was designed to account for intermittent visibility and varying transmission parameters across the free-space medium, and in the quantum node interface, where new submodules were introduced to regulate the flow of packets and prevent nodes from attempting to communicate when the receiving node is not in sight. 

For each free-space link, the link distance, elevation and atmospheric transmission over time
are precomputed (see Fig.\@~\ref{fig:LinkParams}) and stored in \texttt{csv} files, which the \texttt{FreeSpaceChannel} takes as input 
to simulate the varying link over time. This \texttt{csv} file based strategy
is intended to make the interfacing of our simulation to more elaborate orbital dynamics and/or
atmospheric simulation tools easy.

Additionally, the entanglement generation timing procedure was patched to account for differential latency by re-negotiating emission timings at the beginning of every entanglement distribution round.
\section{Conclusion and Outlook}
\label{sec:conclusion}
We have presented a simple analytical model estimating the performance of a quantum satellite link equipped with quantum memory, and applied it to the entanglement distribution rate of a quantum satellite link to one and two ground stations, showing the key importance of quantum memory size in the performance of such high-latency quantum links.
We also developed an extension to a pre-existing simulation platform to enable simulation of satellite links and cross-validated it with our theoretical model.

We plan  to extend our analytical framework in several ways, such as including satellite--to--satellite links for intra- and inter-orbit communication. Moreover, it would be interesting to complement the model with a discussion of the impact of quantum memory noise and channel quality on entanglement distribution. 

Simulation-wise, it could be interesting to further extend QuISP to allow for satellites with dynamic memory allocation: this would not only enable the validation of the performance margin we observed in sec. \ref{sec:TheoryDualLink}, but also pave the way for the study of more refined dynamic memory allocation policies. Finally, since one of the key drivers behind the choice of QuISP was the ease of integration in large networks, we plan to integrate our link in large internetworking setups.

\section*{Acknowledgment}
PF and FG warmly thank Ryosuke Satoh for guiding them through the internals of QuISP, and
Valentina Marulanda Acosta and Matteo Schiavon for introducing them to satellite quantum communications, as well as generating
the orbital data that was used in this work.
PF and FG acknowledge funding by the French state through the \emph{Programme d’Investissements d’Avenir} managed by the Agence Nationale de la Recherche (project ANR-21-CMAQ-0001)
and by the European Union’s \emph{Horizon 2020 research and innovation program} under
Grant Agreement No.\ 820445 and project name
\emph{``Quantum Internet Alliance''}.
PF thanks the Japanese-French Laboratory for Informatics for funding his stay at Keio,
thus making this work possible.
The work of NB, MH and RDV was supported by JST [Moonshot R\&D Program]
Grant Number [JPMJMS226C].

\bibliographystyle{IEEEtran}
\bibliography{IEEEabrv,bibliography}
\end{document}